\documentclass[aps,showpacs,prb,floatfix,preprint,amsmath,superscriptaddress,citeautoscript]{revtex4-1}
\usepackage{graphicx}
\usepackage{amssymb}
\usepackage{bm}
\usepackage{subfigure}
\usepackage{units}
\usepackage{hyperref}
\usepackage[usenames]{color}
\hypersetup{pdfborder=0 0 0,colorlinks=true,citecolor=blue,linkcolor=blue}
\input epsf

\begin{document}

\author{Deniz \c{C}ak{\i}r}
\altaffiliation[Present address: ]{Department of Physics, University of Antwerp, Groenenborgerlaan 171, B-2020 Antwerp, Belgium.}
\affiliation{Computational Materials Science, Faculty of Science and Technology and MESA+ Institute for Nanotechnology, University of Twente,
P.O. Box 217, 7500 AE Enschede, The Netherlands}
\affiliation{Collective Interactions Unit, Okinawa Institute of Science and Technology Graduate University (OIST), Okinawa 904-0495, Japan}
\author{Diana M. Ot\'{a}lvaro}
\affiliation{Computational Materials Science, Faculty of Science and Technology and MESA+ Institute for Nanotechnology, University of Twente,
P.O. Box 217, 7500 AE Enschede, The Netherlands}
\author{Geert Brocks}
\email{g.h.l.a.brocks@utwente.nl}
\affiliation{Computational Materials Science, Faculty of Science and Technology and MESA+ Institute for Nanotechnology, University of Twente,
P.O. Box 217, 7500 AE Enschede, The Netherlands}

\title{From spin-polarized interfaces to giant magnetoresistance in organic spin valves}

\begin{abstract}
We calculate the spin-polarized electronic transport through a molecular bilayer spin valve from first principles, and establish the link between the magnetoresistance and the spin-dependent interactions at the metal-molecule interfaces. The magnetoresistance of a Fe$|$bilayer-C$_{70}|$Fe spin valve attains a high value of 70\% in the linear response regime, but it drops sharply as a function of the applied bias. The current polarization has a value of 80\% in linear response, and also decreases as a function of bias. Both these trends can be modelled in terms of prominent spin-dependent Fe$|$C$_{70}$ interface states close to the Fermi level, unfolding the potential of spinterface science to control and optimize spin currents.
\end{abstract}

\date{\today}
\pacs{72.25.Mk,73.40.Sx,75.47.De,75.78.-n}
\maketitle

\section{Introduction} Carbon-based materials have moved into the focus of spintronics research, because the weak spin-orbit coupling and hyperfine interactions in carbon-based semiconductors generate the prospect of stable spin currents and robust spin operations \cite{Tombros07,Dediu09}. Giant magnetoresistance (MR) effects have been reported in vertical spin valves with layers of organic molecules such as tris(8-hydroxy-quinolinato)-aluminium (Alq$_3$) or fullerenes such as C$_{60}$ \cite{Xiong04,Sun10,Barraud10,Schulz11,Gobbi11,Tran12,Zhang13}. Barraud \textit{ et al.} \cite{Barraud10} have argued that spin-dependent interactions at the interfaces between molecular materials and ferromagnetic electrodes play a pivotal role in the magneto-transport properties of these molecular semiconductor devices. This has prompted the suggestion that highly spin-polarized currents in spintronic devices may be obtained by exploiting such interface interactions, which has been dubbed ``spinterface science'' \cite{Sanvito10}.  

Establishing a direct link between interface properties and spin-dependent transport would be a significant step forward in understanding organic spin valves. Photoemission spectroscopy, scanning tunnelling microscopy (STM), and first-principles calculations enable a detailed analysis of the spin-dependent electronic properties of metal-organic interfaces \cite{Atodiresei10,Brede10,Javaid10,Tran11,Tran13,Djeghloul13}, but a direct connection between these properties and the magneto-transport in organic spin valves is lacking so far. In the field of single molecule electronics, where MR effects have been demonstrated  with STM \cite{Brede10,Schmaus11,Miyamachi12}, first-principles transport calculations have provided detailed descriptions \cite{Rocha05,Ning08,Koleini12}. Two metal electrodes interacting through a single molecule are however generally not a good model for organic devices comprising molecular multilayers.  

In this paper we calculate the spin-dependent current through a prototype spin valve, which consists of a $\sim 2$ nm thick molecular bilayer sandwiched between two ferromagnetic metal electrodes, using a first-principles non-equilibrium Green's function technique. We devise a model where the transmission through a molecular multilayer is factorized, based upon partitioning the system into right and left interface parts,  each consisting of a molecular monolayer adsorbed on a metal surface. This allows for an analysis of the MR and the current polarization in terms of the spin-polarized interface states, both in linear response and at finite bias.

\begin{figure}
\begin{center}
\includegraphics[width=8.5cm]{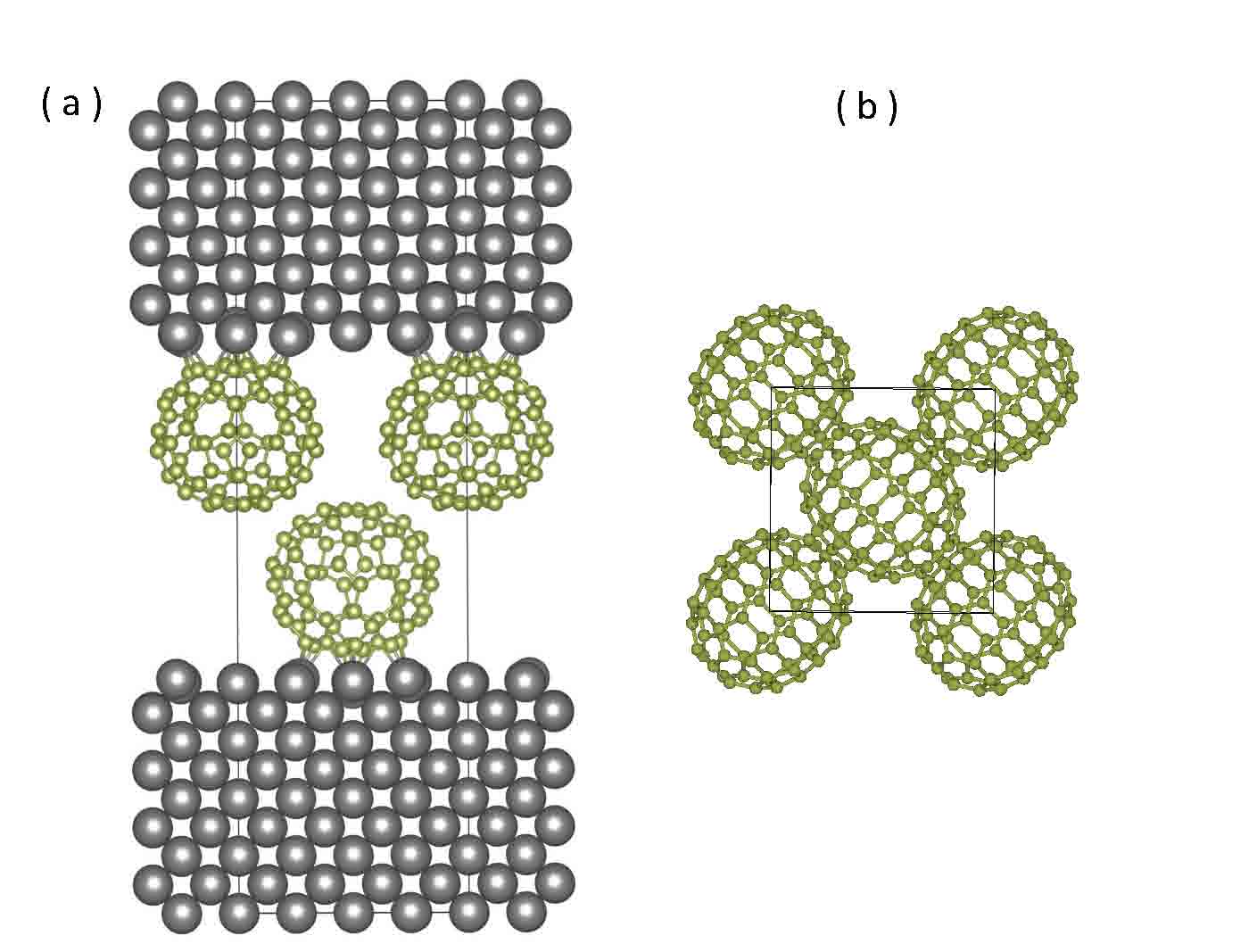}
\caption{(a) Side view of the Fe(001)$|$bilayer-C$_{70}|$Fe(001) structure. (b) Top view with the Fe electrodes removed [\onlinecite{VESTA}.]}
\label{fig:structures}
\end{center}
\end{figure}

We study Fe$|$bilayer-C$_{70}|$Fe spin valves, cf. Fig. \ref{fig:structures}. The bcc-Fe(001) surface is a well-established substrate for organic spintronics that allows for a controlled growth of fullerene layers \cite{Tran11,Tran13}. Fullerene molecules are particularly interesting candidates for applications in spintronics due to the absence of hydrogen atoms that lead to spin de-phasing via hyperfine interactions. In particular, we find that adsorption of C$_{70}$ on Fe(001) results in a favourable interface electronic structure, which gives a large current polarization of 78\% and a large MR of 67\%  \cite{footnote:MR}.

\section{Theory} 
\label{sec:theory}
We start from the Landauer expression for the current at finite bias $V$ and zero temperature \cite{Landauer57}
\begin{equation}
I^\sigma=\frac{e}{h}\int_{E_{F}-\frac{1}{2}eV}^{E_{F}+\frac{1}{2}eV}T^{\sigma}(E,V) dE,\label{eq:1}
\end{equation} 
with $\sigma=\uparrow$$(\downarrow)$ labelling the majority (minority) spin states, and $T^{\sigma}=\mathrm{Tr}\left[\mathbf{\boldsymbol{\Gamma}}_\mathrm{R}^{\sigma}\mathbf{G}_\mathrm{RL}^{\sigma,\mathrm{r}}\boldsymbol{\Gamma}_\mathrm{L}^{\sigma}\boldsymbol{G}_\mathrm{LR}^{\sigma,\mathrm{a}}\right]$ the transmission probability expressed in non-equilibrium Green's functions (NEGF) \cite{Transiesta02}.
$\mathbf{G}_\mathrm{RL}^{\sigma,\mathrm{r(a)}}$ is the retarded (advanced) Green's function
matrix block connecting the right and left leads via the scattering
region, and $\mathbf{\boldsymbol{\Gamma}}^\sigma_\mathrm{R(L)}=-2\mathrm{Im}\boldsymbol{\Sigma}^\sigma_\mathrm{R(L)}$,
with $\boldsymbol{\Sigma}^\sigma_\mathrm{R(L)}$ the self-energy matrix connecting
the scattering region to the right (left) lead \cite{Transiesta02,Khomyakov05}.

Partitioning the Hamiltonian of the scattering region into a left and a right part, one can write\cite{Mingo96,Mathon97}
\begin{equation}
\mathbf{G}^\sigma_\mathrm{RL} = \mathbf{g}^\sigma_\mathrm{R}\mathbf{H}^\sigma_\mathrm{RL}\left(\mathbf{I}_\mathrm{L}-\mathbf{g}^\sigma_\mathrm{L}\mathbf{H}^\sigma_\mathrm{LR}\mathbf{g}^\sigma_\mathrm{R}\mathbf{H}^\sigma_\mathrm{RL}\right)^{-1}\mathbf{g}^\sigma_\mathrm{L},\label{eq:4} 
\end{equation} 
where $\mathbf{g}^\sigma_\mathrm{R(L)}$ is the Green's function matrix of the right
(left) part uncoupled from the left (right) part, and $\mathbf{H}^\sigma_\mathrm{RL}=\left(\mathbf{H}^\sigma_\mathrm{LR}\right)^{\dagger}$
is the Hamilton matrix block that couples the two parts.
Neglecting multiple internal reflections, one can approximate $\mathbf{G}^\sigma_\mathrm{RL}\approx\mathbf{g}^\sigma_\mathrm{R}\mathbf{H}^\sigma_\mathrm{RL}\mathbf{g}^\sigma_\mathrm{L}$. From this approximation and the relation $\mathbf{g}_\mathrm{R(L)}^{\sigma,a}\mathbf{\boldsymbol{\Gamma}}^\sigma_\mathrm{R(L)}\mathbf{g}_\mathrm{R(L)}^{\sigma,r}=2\pi\mathbf{n}^\sigma_\mathrm{R(L)}$, with $\mathbf{\boldsymbol{n}}^\sigma_\mathrm{R(L)}=-\pi^{-1}\mathrm{Im}\mathbf{g}_\mathrm{R(L)}^{\sigma,\mathrm{r}}$
the spectral density matrix of the right (left) part, it then follows\cite{Mingo96,Meunier98}  
\begin{equation}
 T^{\sigma}\approx\mathrm{4\pi^{2}Tr}\left[\mathbf{\boldsymbol{n}}_\mathrm{R}^{\sigma}\mathbf{H}_\mathrm{RL}^{\sigma}\boldsymbol{n}_\mathrm{L}^{\sigma}\mathbf{H}_\mathrm{LR}^{\sigma}\right].
\end{equation} 

In a representation where the spectral density matrix is diagonal, one of the matrix elements is much larger than the other ones, if a single molecular state is dominant (depending on the energy, the HOMO or LUMO, for instance). Setting all but one matrix element to zero in the density matrices of the left and the right parts, the transmission can be approximated by $T^{\sigma}\approx4\pi^{2}|H^\sigma|^{2}n_\mathrm{R}^{\sigma}n_\mathrm{L}^{\sigma}$,
with $n^\sigma_\mathrm{R(L)}$ the projected density of states (PDOS), i.e., projected on the molecules at the right (left) electrode.  Using this expression in Eq. (\ref{eq:1}) in linear response ($V\rightarrow0$) leads to the Julli\`{e}re expression for the MR \cite{Julliere75}. In the original Julli\`{e}re model, bulk DOSs of the ferromagnetic electrodes are used to calculate the MR. It is more appropriate to use interface DOSs, but the local DOS in a metal$|$insulator$|$metal junction gradually changes from the metal into the insulator, making it difficult to pinpoint an exact interface DOS \cite{Moodera99}. For a metal-molecule interface the PDOS $n^\sigma_\mathrm{R(L)}$ provides a unique interface DOS. 

Expressing the transmission $T^{\sigma}$ in terms of a product of PDOSs of the right and left interface, means that the transmission through an asymmetric system, where right and left interfaces are different from one another, can be approximated by a geometrical average  $T^{\sigma}=\sqrt{T_\mathrm{R}^{\sigma}}\sqrt{T_\mathrm{L}^{\sigma}}$ \cite{Belashchenko04,Xu06}. Here $T_\mathrm{R(L)}^{\sigma}$ is the transmission through a symmetric system with identical right and left interfaces, i.e., characterized by the same PDOS, so $\sqrt{T_\mathrm{R(L)}^{\sigma}} \propto n^\sigma_\mathrm{R(L)}$. If in addition we assume that the PDOSs are not affected by the bias $V$ except for a rigid shift, then similar factorization arguments lead to the expressions
\begin{equation}
T_\mathrm{P}^{\sigma}\left(E,V\right)=\sqrt{T_\mathrm{P}^{\sigma}\left(E-\frac{eV}{2},0\right)}\sqrt{T_\mathrm{P}^{\sigma}\left(E+\frac{eV}{2},0\right)},\label{eq:11}
\end{equation}
\begin{equation}
T_\mathrm{AP}^{\sigma}\left(E,V\right)=\sqrt{T_\mathrm{P}^{\sigma}\left(E-\frac{eV}{2},0\right)}\sqrt{T_\mathrm{P}^{-\sigma}\left(E+\frac{eV}{2},0\right)},\label{eq:10}
\end{equation}
where P (AP) describes the situation with the magnetizations of the two ferromagnetic electrodes parallel (anti-parallel). With these approximations one can construct the P transmission spectrum at finite bias, or the AP transmission at any bias, starting from the P spectrum at zero bias, which greatly facilitates the interpretation of the MR effect and of the $I$-$V$ curves.
 
\section{Computational details.} We optimize the structure of the Fe(001)$|$C$_{70}$ interface, using density functional theory (DFT) at the spin-polarized generalized gradient approximation (GGA/PBE) level, as implemented in VASP\cite{vasp-1,vasp-2}. The same computational parameters are used as in Ref. \onlinecite{Tran13}. The interface is modeled using a 4$\times$4 Fe(001) surface unit cell (cell parameter 11.32 \AA), containing one C$_{70}$ molecule. 
The distance between nearest neighbor molecules is then slightly larger than the 10-11 \AA\ observed in the C$_{70}$ molecular crystal.\cite{Verheijen:cp92} 
A slab of seven atomic layers represents the Fe(001) substrate. The molecules and the uppermost three Fe atomic layers are relaxed. A structure for the bilayer junction, Fe$|$C$_{70}$-C$_{70}|$Fe, is generated by mirroring the optimized Fe(001)$|$C$_{70}$ structure, rotating the mirror image by 90$^\mathrm{o}$, and translating it in plane by half a cell, see Fig. \ref{fig:structures}. The spacing between the C$_{70}$ layers is such that the shortest intermolecular C--C distance is 3.2 \AA, which is a typical value for close-packed fullerenes or carbon nanotubes. 

Electronic transport in the bilayer junction is calculated using the self-consistent NEGF technique as implemented in TranSIESTA \cite{siesta,Transiesta02}. Single-$\zeta$ and double-$\zeta$ (plus polarization) numerical orbital basis sets are used for Fe and C, respectively. We employ the GGA/PBE functional and norm-conserving pseudo-potentials \cite{tm,footnote:SIESTA}. A 6$\times$6 in-plane $k$-point mesh is adequate to obtain sufficiently accurate transport results. For instance, the total conductance at small bias is then converged on a scale of 2\%.

\section{Results}
\subsection{Fe$|$C$_{70}$ interface} 
From a number of possible adsorption geometries, we have identified a structure as most stable where the long axis of the C$_{70}$ molecules is parallel to the surface. Two neighbouring C$_{70}$ hexagons are nearly parallel to the surface and the edge shared by these two hexagons is on top of a surface Fe atom. The shortest Fe--C bonds are in the range 2.0-2.3 \AA, which is indicative of a strong (chemisorption) interaction between C$_{70}$ and the Fe substrate, as confirmed by the calculated binding energy of 3.0 eV. Nevertheless, the geometry of the C$_{70}$ molecule is only mildly affected by the adsorption.

\begin{figure}
\begin{center}
\includegraphics[width=8.5cm]{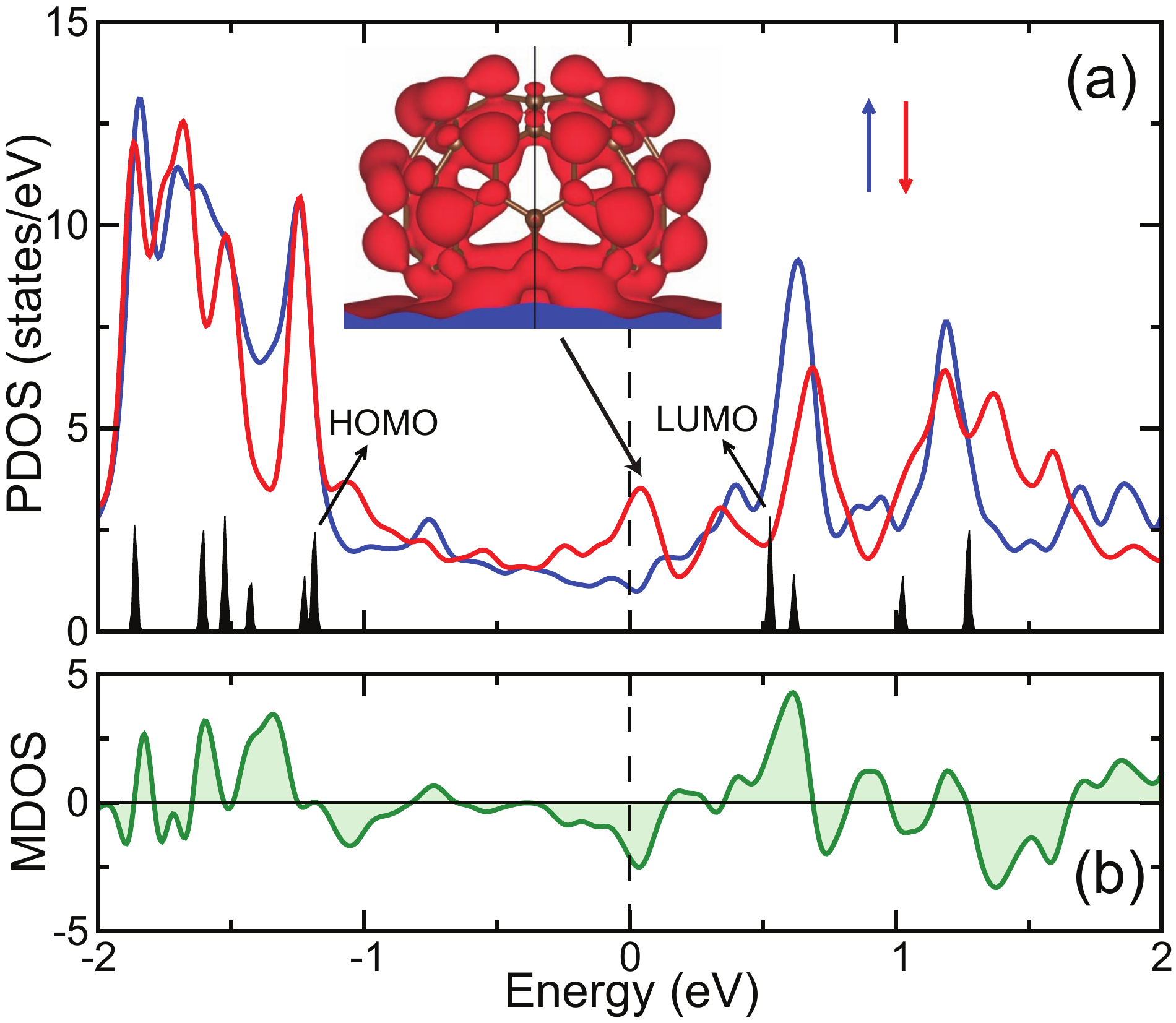}
\caption{(a) P(rojected)DOS $n^\uparrow$ of majority (blue) and $n^\downarrow$ of minority (red) spin states of the Fe(001)$|$C$_{70}$ interface, summed over the C$_{70}$ atoms, as calculated with VASP. Gaussian smearing with a smearing parameter of 0.05 eV is applied. The zero of energy is at the Fermi level $E_\mathrm{F}$. The black lines indicate the DFT energy levels of the isolated C$_{70}$ molecule. (b) M(agnetization)DOS $\Delta n=n^\uparrow-n^\downarrow$, in states/eV. The inset to (a) shows the local MDOS at the Fermi level, illustrating its minority spin character.  }
\label{fig:DOS}
\end{center}
\end{figure}

Figure \ref{fig:DOS} shows the PDOS summed over all atoms of the molecule. The DFT levels of an isolated C$_{70}$ molecule are given for comparison, aligned with the PDOS through the lowest $\sigma$-levels, which are unperturbed by adsorption. In contrast, adsorption significantly broadens and shifts the molecular $\pi$-states, due to hybridization with the substrate. Despite the large perturbation, it is still possible to assign molecular labels to the peaks in the PDOS. The peaks at $-1.2$ eV and $+0.6$ eV with respect to $E_\mathrm{F}$ have molecular HOMO and LUMO character, respectively, and the peak at $E_\mathrm{F}$ in the minority spin states also has LUMO character. 

The spin-polarized states of the substrate interact differently with the molecule, resulting in a markedly different PDOS for the two spin states. Around the Fermi level the interaction with the minority spin states is particularly strong, consistent with the fact that the Fe(001) surface has prominent minority spin surface resonances in this energy range \cite{Stroscio95}. The interaction between molecule and surface induces a magnetic moment of 0.26 $\mu_B$ on the C$_{70}$ molecule in the minority spin direction, which is similar to the induced moment ofn C$_{60}$ on Fe(001) \cite{Tran13}.

The difference between the PDOSs of the two spin states gives a magnetization density of states (MDOS) $\Delta n(E)=n^\uparrow(E)-n^\downarrow(E)$, shown in Fig. \ref{fig:DOS}(b). A MDOS that oscillates similarly as a function of the energy has been observed at the C$_{60}|$Fe(001) interface \cite{Tran11}. For transport the energy region around the Fermi level is most relevant, where the MDOS has a (negative) peak. This peak gives a spin polarization $\Delta n/(n^\uparrow+n^\downarrow) \approx -40$\% at $E = E_\mathrm{F}$, which according to the Julli{\`{e}}re model then gives a MR $\approx 40$\%.

\begin{figure}
\begin{center}
\includegraphics[width=8.5cm]{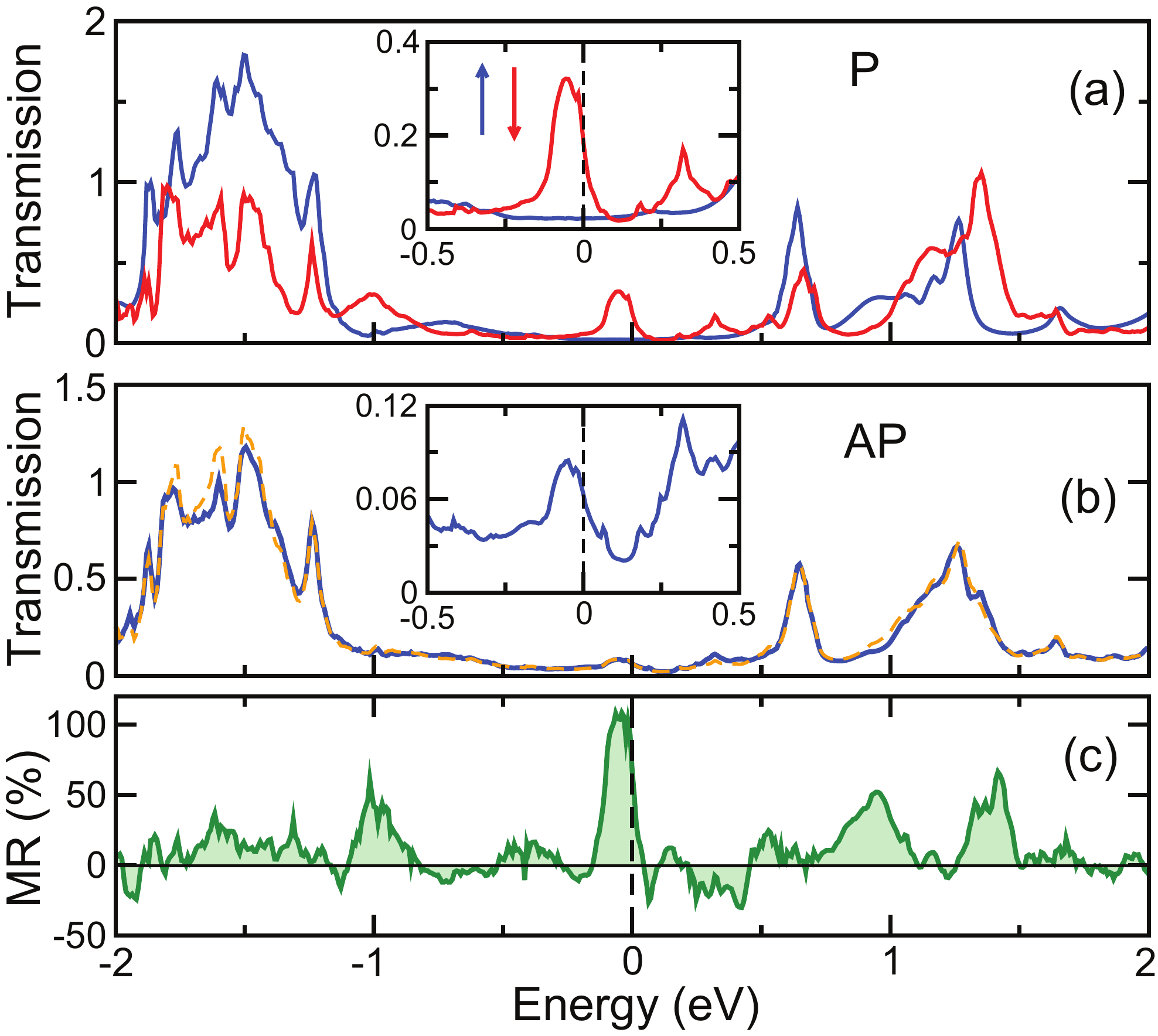}
\caption{(a) Transmissions $T_\mathrm{P}^\uparrow$ of majority (blue) and $T_\mathrm{P}^\downarrow$ of minority (red) spin channels of Fe$|$C$_{70}$-C$_{70}|$Fe at zero bias, as calculated with TranSIESTA. The zero of energy is at the Fermi level $E_\mathrm{F}$. (b) Transmissions $T_\mathrm{AP}^\uparrow = T_\mathrm{AP}^\downarrow$ (blue). The yellow dashed line represents the factorization approximation of  Eq. (\ref{eq:10}). (c) The MR spectrum as a function of energy.}
\label{fig:transmission}
\end{center}
\end{figure}

\subsection{Fe$|$C$_{70}$-C$_{70}|$Fe, linear response}
Figure \ref{fig:transmission} (a) shows the transmission spectra $T_\mathrm{P}^\sigma(E,V=0)$ at zero bias, calculated for the bilayer junction Fe$|$C$_{70}$-C$_{70}|$Fe with the magnetizations of both Fe electrodes parallel (P). The peaks in the transmission correspond to those observed in the PDOS, see Fig. \ref{fig:DOS}, wich suggests that the factorization approximation discussed in Sec. \ref{sec:theory} may be applied. Of particular interest is the peak around the Fermi energy in the minority spin channel, as at low bias this peak dominates the conductance. The corresponding state has substantial LUMO character, and is delocalized over the whole molecule, so that the bilayer C$_{70}$ junction presents a relatively thin barrier. The conductance polarization, defined as $(T^\uparrow - T^\downarrow)/(T^\uparrow + T^\downarrow)$, is $-78$\% at $E=E_\mathrm{F}$ and $V=0$, which is also the value of the current polarization $\mathrm{CP}=(I^\downarrow - I^\uparrow)/(I^\uparrow + I^\downarrow)$ in the linear response regime. The current has a remarkably large spin-polarization, and it is negative because the minority spin dominates.

Figure \ref{fig:transmission} (b) shows the transmission spectra at zero bias, calculated for the bilayer junction with the magnetizations of both Fe electrodes anti-parallel (AP). The factorization approximation of Eq. (\ref{eq:10}) implies that the transmission in the AP case can be constructed as a geometric average of the transmission of the two spin channels in the P case. Figure \ref{fig:transmission} (b) demonstrates that this approximation works very well. The MR in the linear response regime can be calculated replacing the currents $I$ by the corresponding transmissions $T$ at $E=E_\mathrm{F}$ and $V=0$.  From the calculated $T_\mathrm{P(AP)}$ the MR is 67\%, and from the factorization approximation, the MR is 70\%. 

From the PDOSs and the  Julli{\`{e}}re model we obtained a smaller MR of 40\%. One should note however that the MR is very sensitive to the shapes and positions of the peaks in the transmission spectra. Figure \ref{fig:transmission} (c) shows the MR as a function of the energy, calculated from the transmission spectra. The position of the Fermi level is in a narrow peak of the MR spectrum. The maximum of this peak is $\sim110$\% at $E_\mathrm{F}-0.04$ eV.   

\begin{figure}
\begin{center}
\includegraphics[width=8.5cm]{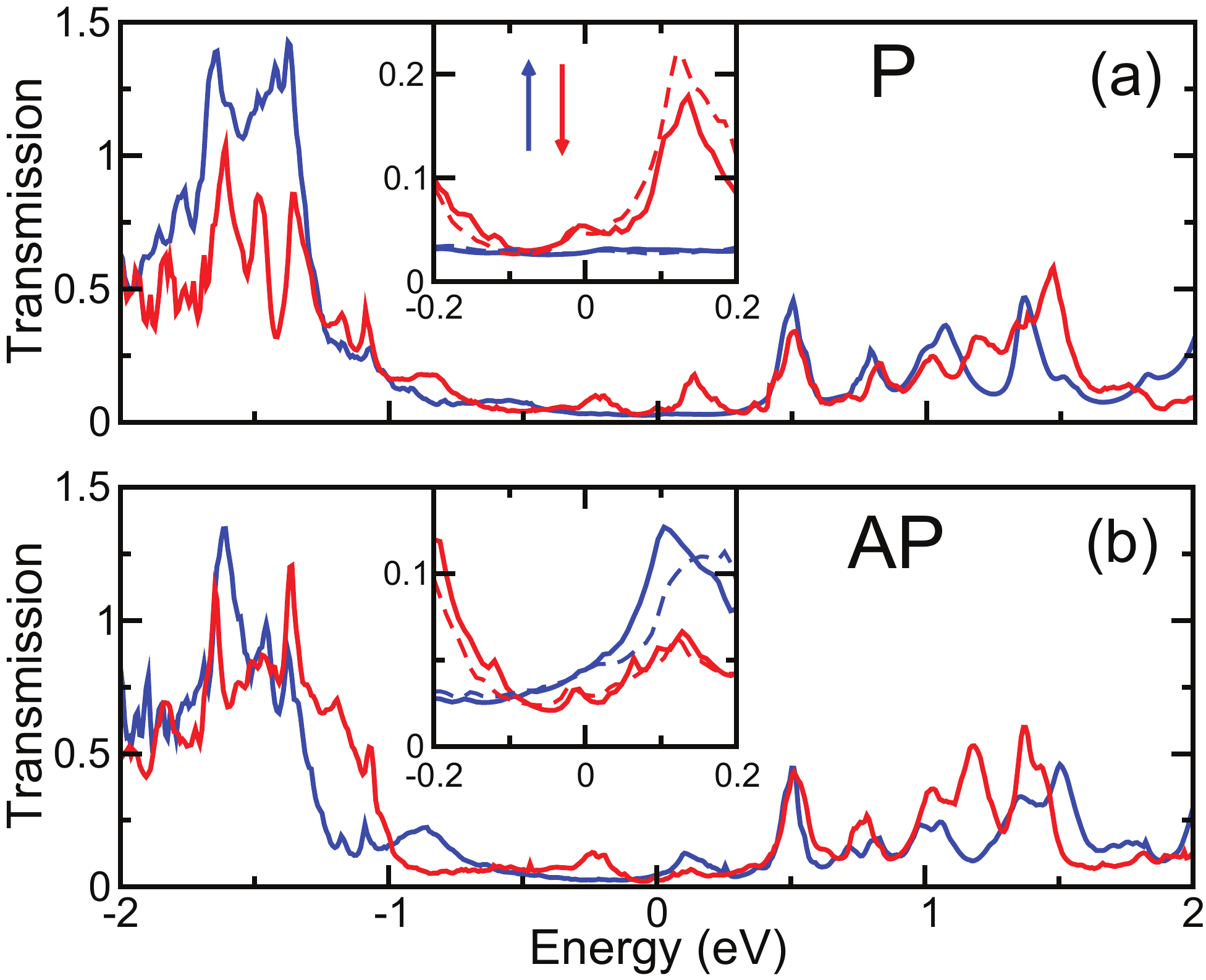}
\caption{(a) Transmissions $T_\mathrm{P}^\uparrow$ of majority (blue) and $T_\mathrm{P}^\downarrow$ of minority (red) spins of Fe$|$C$_{70}$-C$_{70}|$Fe at bias $V=0.4$V. (b)  $T_\mathrm{AP}^\uparrow$ of majority (blue) and $T_\mathrm{AP}^\downarrow$ of minority (red) spins. The dashed lines indicate the factorization approximations of Eqs. (\ref{eq:11}) and (\ref{eq:10}).}
\label{fig:finitebias}
\end{center}
\end{figure}

\subsection{Fe$|$C$_{70}$-C$_{70}|$Fe, finite bias}
Figure \ref{fig:finitebias} shows transmission spectra $T_\mathrm{P}^\sigma(E,V)$ at a bias $V=0.4$ V, calculated self-consistently. To obtain the current, Eq. (\ref{eq:1}), one has to integrate the transmission from $E=-0.2$ to $E=0.2$ eV, see the insets of Fig. \ref{fig:finitebias}. The currents for the P and AP cases become very similar, resulting in a small MR. The transmission can be interpreted starting from the zero bias transmission using Eqs. (\ref{eq:11}) and (\ref{eq:10}). $T_\mathrm{P}^\sigma(E,V=0)$ has a prominent peak in the minority spin channel close to $E_\mathrm{F}$ corresponding to the LUMO derived state at $E_\mathrm{F}$, cf. Figs. \ref{fig:DOS}(a) and \ref{fig:transmission}(a). Factorization according to Eqs. (\ref{eq:11}) and (\ref{eq:10}) splits this peak and shifts the factors by $\pm eV/2$, such that two peaks appear at $E_\mathrm{F} \pm eV/2$, respectively. This construction is shown as the dotted lines in Fig. \ref{fig:finitebias}. For the P case, Eq. (\ref{eq:11}), both these peaks appear in the minority spin channel $T_\mathrm{P}^\downarrow$. The CP should therefore be still significant at finite bias (albeit smaller than at zero bias). For the AP case at finite bias, Eq. (\ref{eq:10}), one peak appears in  $T_\mathrm{AP}^\downarrow$ and  the other in $T_\mathrm{AP}^\uparrow$. As we integrate over these peaks, the MR at finite bias should therefore be much smaller than at zero bias. One expects that the MR drops sharply with increasing bias, as the peak in the minority spin channel moves away from the Fermi level.

\begin{figure}
\begin{center}
\includegraphics[width=8.5cm]{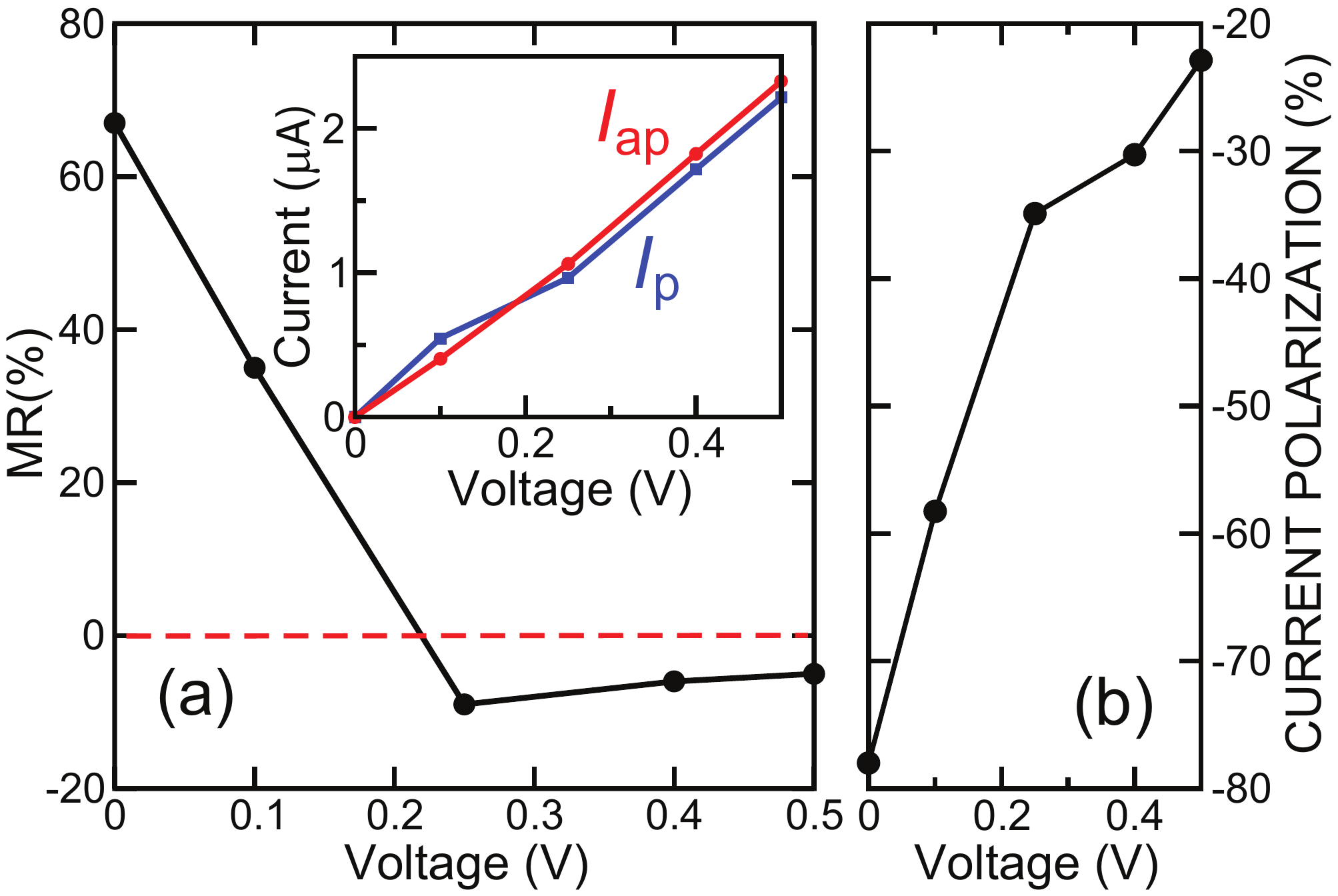}
\caption{(a) Magnetoresistance (MR) as function of the bias $V$. The inset show the total currents $I_\mathrm{P}$ and  $I_\mathrm{AP}$ as function of $V$. (b) Current polarization (CP) of $I_\mathrm{P}$  as function of $V$. }
\label{fig:MR}
\end{center}
\end{figure}

This is confirmed by the self-consistent calculations shown in Fig. \ref{fig:MR}(a). At a bias $V=0.1$ V the MR is roughly halved, and it reaches small (negative) values, $-10$\% $<$ MR $< 0$\%, for biases $V\geq0.25$ V. A similar drop of the MR as function of bias has been observed in Alq$_3$ tunnel barriers \cite{Barraud10}. Because of the delocalized nature of the hybridized Fe(001)$|$C$_{70}$ interface states, a bilayer of C$_{70}$ molecules is still quite transparent, however, which means that the currents do not show the exponential dependence on bias that is characteristic of tunnel barriers. The absolute value of the CP decreases monotonically with increasing applied bias, see Fig. \ref{fig:MR}(b), which agrees with the analysis given in the previous paragraph. 
 
\section{Summary} We calculate from first principles the spin-polarized transport in Fe$|$bilayer-C$_{70}|$Fe devices as a function of applied bias. We show that transport in such organic spin valves can be analyzed with a factorization model, which enables us to interpret the transmission in terms of the Fe$|$C$_{70}$ interface states. This opens a route toward exploiting spin-dependent metal-molecule interactions to optimize spin currents. In particular we show that adsorption of C$_{70}$ on Fe(001) results in a sizeable spin-polarization at the Fermi level. The current spin-polarization has a maximum value of 78\% in the linear response regime, and it decreases monotonically as function of the applied bias. The magnetoresistance has a value of $\sim 67$\% at linear response, and it decreases rapidly with the applied bias.

\begin{acknowledgements}
We thank Michel de Jong and Zhe Yuan for useful discussions. 
Computational resources were provided through the Physical Sciences division of the Netherlands Organization for Scientific Research (NWO-EW) and by TUBITAK ULAKBIM, High Performance and Grid Computing Center (TR-Grid e-Infrastructure). 
\end{acknowledgements}                    
                                                                                                                               

\begin{thebibliography}{40}%
\makeatletter
\providecommand \@ifxundefined [1]{%
 \@ifx{#1\undefined}
}%
\providecommand \@ifnum [1]{%
 \ifnum #1\expandafter \@firstoftwo
 \else \expandafter \@secondoftwo
 \fi
}%
\providecommand \@ifx [1]{%
 \ifx #1\expandafter \@firstoftwo
 \else \expandafter \@secondoftwo
 \fi
}%
\providecommand \natexlab [1]{#1}%
\providecommand \enquote  [1]{``#1''}%
\providecommand \bibnamefont  [1]{#1}%
\providecommand \bibfnamefont [1]{#1}%
\providecommand \citenamefont [1]{#1}%
\providecommand \href@noop [0]{\@secondoftwo}%
\providecommand \href [0]{\begingroup \@sanitize@url \@href}%
\providecommand \@href[1]{\@@startlink{#1}\@@href}%
\providecommand \@@href[1]{\endgroup#1\@@endlink}%
\providecommand \@sanitize@url [0]{\catcode `\\12\catcode `\$12\catcode
  `\&12\catcode `\#12\catcode `\^12\catcode `\_12\catcode `\%12\relax}%
\providecommand \@@startlink[1]{}%
\providecommand \@@endlink[0]{}%
\providecommand \url  [0]{\begingroup\@sanitize@url \@url }%
\providecommand \@url [1]{\endgroup\@href {#1}{\urlprefix }}%
\providecommand \urlprefix  [0]{URL }%
\providecommand \Eprint [0]{\href }%
\providecommand \doibase [0]{http://dx.doi.org/}%
\providecommand \selectlanguage [0]{\@gobble}%
\providecommand \bibinfo  [0]{\@secondoftwo}%
\providecommand \bibfield  [0]{\@secondoftwo}%
\providecommand \translation [1]{[#1]}%
\providecommand \BibitemOpen [0]{}%
\providecommand \bibitemStop [0]{}%
\providecommand \bibitemNoStop [0]{.\EOS\space}%
\providecommand \EOS [0]{\spacefactor3000\relax}%
\providecommand \BibitemShut  [1]{\csname bibitem#1\endcsname}%
\let\auto@bib@innerbib\@empty
\bibitem [{\citenamefont {Tombros}\ \emph {et~al.}(2007)\citenamefont
  {Tombros}, \citenamefont {Jozsa}, \citenamefont {Popinciuc}, \citenamefont
  {Jonkman},\ and\ \citenamefont {van Wees}}]{Tombros07}%
  \BibitemOpen
  \bibfield  {author} {\bibinfo {author} {\bibfnamefont {N.}~\bibnamefont
  {Tombros}}, \bibinfo {author} {\bibfnamefont {C.}~\bibnamefont {Jozsa}},
  \bibinfo {author} {\bibfnamefont {M.}~\bibnamefont {Popinciuc}}, \bibinfo
  {author} {\bibfnamefont {H.~T.}\ \bibnamefont {Jonkman}}, \ and\ \bibinfo
  {author} {\bibfnamefont {B.~J.}\ \bibnamefont {van Wees}},\ }\href@noop {}
  {\bibfield  {journal} {\bibinfo  {journal} {Nature}\ }\textbf {\bibinfo
  {volume} {448}},\ \bibinfo {pages} {571} (\bibinfo {year}
  {2007})}\BibitemShut {NoStop}%
\bibitem [{\citenamefont {Dediu}\ \emph {et~al.}(2009)\citenamefont {Dediu},
  \citenamefont {Hueso}, \citenamefont {Bergenti},\ and\ \citenamefont
  {Taliani}}]{Dediu09}%
  \BibitemOpen
  \bibfield  {author} {\bibinfo {author} {\bibfnamefont {V.~A.}\ \bibnamefont
  {Dediu}}, \bibinfo {author} {\bibfnamefont {L.~E.}\ \bibnamefont {Hueso}},
  \bibinfo {author} {\bibfnamefont {I.}~\bibnamefont {Bergenti}}, \ and\
  \bibinfo {author} {\bibfnamefont {C.}~\bibnamefont {Taliani}},\ }\href@noop
  {} {\bibfield  {journal} {\bibinfo  {journal} {Nature Mater.}\ }\textbf
  {\bibinfo {volume} {8}},\ \bibinfo {pages} {707} (\bibinfo {year}
  {2009})}\BibitemShut {NoStop}%
\bibitem [{\citenamefont {Xiong}\ \emph {et~al.}(2004)\citenamefont {Xiong},
  \citenamefont {Wu},\ and\ \citenamefont {Vardeny}}]{Xiong04}%
  \BibitemOpen
  \bibfield  {author} {\bibinfo {author} {\bibfnamefont {Z.~H.}\ \bibnamefont
  {Xiong}}, \bibinfo {author} {\bibfnamefont {D.}~\bibnamefont {Wu}}, \ and\
  \bibinfo {author} {\bibfnamefont {V.}~\bibnamefont {Vardeny}},\ }\href@noop
  {} {\bibfield  {journal} {\bibinfo  {journal} {Nature}\ }\textbf {\bibinfo
  {volume} {427}},\ \bibinfo {pages} {821} (\bibinfo {year}
  {2004})}\BibitemShut {NoStop}%
\bibitem [{\citenamefont {Sun}\ \emph {et~al.}(2010)\citenamefont {Sun},
  \citenamefont {Yin}, \citenamefont {Sun}, \citenamefont {Guo}, \citenamefont
  {Gai}, \citenamefont {Zhang}, \citenamefont {Ward}, \citenamefont {Cheng},\
  and\ \citenamefont {Shen}}]{Sun10}%
  \BibitemOpen
  \bibfield  {author} {\bibinfo {author} {\bibfnamefont {D.}~\bibnamefont
  {Sun}}, \bibinfo {author} {\bibfnamefont {L.}~\bibnamefont {Yin}}, \bibinfo
  {author} {\bibfnamefont {C.}~\bibnamefont {Sun}}, \bibinfo {author}
  {\bibfnamefont {H.}~\bibnamefont {Guo}}, \bibinfo {author} {\bibfnamefont
  {Z.}~\bibnamefont {Gai}}, \bibinfo {author} {\bibfnamefont {X.-G.}\
  \bibnamefont {Zhang}}, \bibinfo {author} {\bibfnamefont {T.~Z.}\ \bibnamefont
  {Ward}}, \bibinfo {author} {\bibfnamefont {Z.}~\bibnamefont {Cheng}}, \ and\
  \bibinfo {author} {\bibfnamefont {J.}~\bibnamefont {Shen}},\ }\href@noop {}
  {\bibfield  {journal} {\bibinfo  {journal} {Phys. Rev. Lett.}\ }\textbf
  {\bibinfo {volume} {104}},\ \bibinfo {pages} {236602} (\bibinfo {year}
  {2010})}\BibitemShut {NoStop}%
\bibitem [{\citenamefont {Barraud}\ \emph {et~al.}(2010)\citenamefont
  {Barraud}, \citenamefont {Seneor}, \citenamefont {Mattana}, \citenamefont
  {Fusil}, \citenamefont {Bouzehouane}, \citenamefont {Deranlot}, \citenamefont
  {Graziosi}, \citenamefont {Hueso}, \citenamefont {Bergenti}, \citenamefont
  {Dediu}, \citenamefont {Petroff},\ and\ \citenamefont {Fert}}]{Barraud10}%
  \BibitemOpen
  \bibfield  {author} {\bibinfo {author} {\bibfnamefont {C.}~\bibnamefont
  {Barraud}}, \bibinfo {author} {\bibfnamefont {P.}~\bibnamefont {Seneor}},
  \bibinfo {author} {\bibfnamefont {R.}~\bibnamefont {Mattana}}, \bibinfo
  {author} {\bibfnamefont {S.}~\bibnamefont {Fusil}}, \bibinfo {author}
  {\bibfnamefont {K.}~\bibnamefont {Bouzehouane}}, \bibinfo {author}
  {\bibfnamefont {C.}~\bibnamefont {Deranlot}}, \bibinfo {author}
  {\bibfnamefont {P.}~\bibnamefont {Graziosi}}, \bibinfo {author}
  {\bibfnamefont {L.}~\bibnamefont {Hueso}}, \bibinfo {author} {\bibfnamefont
  {I.}~\bibnamefont {Bergenti}}, \bibinfo {author} {\bibfnamefont {V.~A.}\
  \bibnamefont {Dediu}}, \bibinfo {author} {\bibfnamefont {F.}~\bibnamefont
  {Petroff}}, \ and\ \bibinfo {author} {\bibfnamefont {A.}~\bibnamefont
  {Fert}},\ }\href@noop {} {\bibfield  {journal} {\bibinfo  {journal} {Nature
  Phys.}\ }\textbf {\bibinfo {volume} {6}},\ \bibinfo {pages} {615} (\bibinfo
  {year} {2010})}\BibitemShut {NoStop}%
\bibitem [{\citenamefont {Schulz}\ \emph {et~al.}(2011)\citenamefont {Schulz},
  \citenamefont {Nuccio}, \citenamefont {Willis}, \citenamefont {Desai},
  \citenamefont {Shakya}, \citenamefont {Kreouzis}, \citenamefont {Malik},
  \citenamefont {Bernhard}, \citenamefont {Pratt}, \citenamefont {Morley},
  \citenamefont {Suter}, \citenamefont {Nieuwenhuys}, \citenamefont {Prokscha},
  \citenamefont {Morenzoni}, \citenamefont {Gillin},\ and\ \citenamefont
  {Drew}}]{Schulz11}%
  \BibitemOpen
  \bibfield  {author} {\bibinfo {author} {\bibfnamefont {L.}~\bibnamefont
  {Schulz}}, \bibinfo {author} {\bibfnamefont {L.}~\bibnamefont {Nuccio}},
  \bibinfo {author} {\bibfnamefont {M.}~\bibnamefont {Willis}}, \bibinfo
  {author} {\bibfnamefont {P.}~\bibnamefont {Desai}}, \bibinfo {author}
  {\bibfnamefont {P.}~\bibnamefont {Shakya}}, \bibinfo {author} {\bibfnamefont
  {T.}~\bibnamefont {Kreouzis}}, \bibinfo {author} {\bibfnamefont {V.~K.}\
  \bibnamefont {Malik}}, \bibinfo {author} {\bibfnamefont {C.}~\bibnamefont
  {Bernhard}}, \bibinfo {author} {\bibfnamefont {F.~L.}\ \bibnamefont {Pratt}},
  \bibinfo {author} {\bibfnamefont {N.~A.}\ \bibnamefont {Morley}}, \bibinfo
  {author} {\bibfnamefont {A.}~\bibnamefont {Suter}}, \bibinfo {author}
  {\bibfnamefont {G.~J.}\ \bibnamefont {Nieuwenhuys}}, \bibinfo {author}
  {\bibfnamefont {T.}~\bibnamefont {Prokscha}}, \bibinfo {author}
  {\bibfnamefont {E.}~\bibnamefont {Morenzoni}}, \bibinfo {author}
  {\bibfnamefont {W.~P.}\ \bibnamefont {Gillin}}, \ and\ \bibinfo {author}
  {\bibfnamefont {A.~J.}\ \bibnamefont {Drew}},\ }\href@noop {} {\bibfield
  {journal} {\bibinfo  {journal} {Nature Mater.}\ }\textbf {\bibinfo {volume}
  {10}},\ \bibinfo {pages} {39} (\bibinfo {year} {2011})}\BibitemShut {NoStop}%
\bibitem [{\citenamefont {Gobbi}\ \emph {et~al.}(2011)\citenamefont {Gobbi},
  \citenamefont {Golmar}, \citenamefont {Llopis}, \citenamefont {Casanova},\
  and\ \citenamefont {Hueso}}]{Gobbi11}%
  \BibitemOpen
  \bibfield  {author} {\bibinfo {author} {\bibfnamefont {M.}~\bibnamefont
  {Gobbi}}, \bibinfo {author} {\bibfnamefont {F.}~\bibnamefont {Golmar}},
  \bibinfo {author} {\bibfnamefont {R.}~\bibnamefont {Llopis}}, \bibinfo
  {author} {\bibfnamefont {F.}~\bibnamefont {Casanova}}, \ and\ \bibinfo
  {author} {\bibfnamefont {L.~E.}\ \bibnamefont {Hueso}},\ }\href@noop {}
  {\bibfield  {journal} {\bibinfo  {journal} {Adv. Mater.}\ }\textbf {\bibinfo
  {volume} {23}},\ \bibinfo {pages} {1609} (\bibinfo {year}
  {2011})}\BibitemShut {NoStop}%
\bibitem [{\citenamefont {Tran}\ \emph {et~al.}(2012)\citenamefont {Tran},
  \citenamefont {Le}, \citenamefont {Sanderink}, \citenamefont {van~der Wiel},\
  and\ \citenamefont {de~Jong}}]{Tran12}%
  \BibitemOpen
  \bibfield  {author} {\bibinfo {author} {\bibfnamefont {T.~L.~A.}\
  \bibnamefont {Tran}}, \bibinfo {author} {\bibfnamefont {T.~Q.}\ \bibnamefont
  {Le}}, \bibinfo {author} {\bibfnamefont {J.~G.~M.}\ \bibnamefont
  {Sanderink}}, \bibinfo {author} {\bibfnamefont {W.~G.}\ \bibnamefont {van~der
  Wiel}}, \ and\ \bibinfo {author} {\bibfnamefont {M.~P.}\ \bibnamefont
  {de~Jong}},\ }\href@noop {} {\bibfield  {journal} {\bibinfo  {journal} {Adv.
  Func. Mater.}\ }\textbf {\bibinfo {volume} {22}},\ \bibinfo {pages} {1180}
  (\bibinfo {year} {2012})}\BibitemShut {NoStop}%
\bibitem [{\citenamefont {Zhang}\ \emph {et~al.}(2013)\citenamefont {Zhang},
  \citenamefont {Mizukami}, \citenamefont {Kubota}, \citenamefont {Ma},
  \citenamefont {Oogane}, \citenamefont {Naganuma}, \citenamefont {Ando},\ and\
  \citenamefont {Miyazaki}}]{Zhang13}%
  \BibitemOpen
  \bibfield  {author} {\bibinfo {author} {\bibfnamefont {X.}~\bibnamefont
  {Zhang}}, \bibinfo {author} {\bibfnamefont {S.}~\bibnamefont {Mizukami}},
  \bibinfo {author} {\bibfnamefont {T.}~\bibnamefont {Kubota}}, \bibinfo
  {author} {\bibfnamefont {Q.}~\bibnamefont {Ma}}, \bibinfo {author}
  {\bibfnamefont {M.}~\bibnamefont {Oogane}}, \bibinfo {author} {\bibfnamefont
  {H.}~\bibnamefont {Naganuma}}, \bibinfo {author} {\bibfnamefont
  {Y.}~\bibnamefont {Ando}}, \ and\ \bibinfo {author} {\bibfnamefont
  {T.}~\bibnamefont {Miyazaki}},\ }\href@noop {} {\bibfield  {journal}
  {\bibinfo  {journal} {Nature Commun.}\ }\textbf {\bibinfo {volume} {4}},\
  \bibinfo {pages} {1392} (\bibinfo {year} {2013})}\BibitemShut {NoStop}%
\bibitem [{\citenamefont {Sanvito}(2010)}]{Sanvito10}%
  \BibitemOpen
  \bibfield  {author} {\bibinfo {author} {\bibfnamefont {S.}~\bibnamefont
  {Sanvito}},\ }\href@noop {} {\bibfield  {journal} {\bibinfo  {journal}
  {Nature Phys.}\ }\textbf {\bibinfo {volume} {6}},\ \bibinfo {pages} {562}
  (\bibinfo {year} {2010})}\BibitemShut {NoStop}%
\bibitem [{\citenamefont {Atodiresei}\ \emph {et~al.}(2010)\citenamefont
  {Atodiresei}, \citenamefont {Brede}, \citenamefont {Lazi\'{c}}, \citenamefont
  {Caciuc}, \citenamefont {Hoffmann}, \citenamefont {Wiesendanger},\ and\
  \citenamefont {Bl\"{u}gel}}]{Atodiresei10}%
  \BibitemOpen
  \bibfield  {author} {\bibinfo {author} {\bibfnamefont {N.}~\bibnamefont
  {Atodiresei}}, \bibinfo {author} {\bibfnamefont {J.}~\bibnamefont {Brede}},
  \bibinfo {author} {\bibfnamefont {P.}~\bibnamefont {Lazi\'{c}}}, \bibinfo
  {author} {\bibfnamefont {V.}~\bibnamefont {Caciuc}}, \bibinfo {author}
  {\bibfnamefont {G.}~\bibnamefont {Hoffmann}}, \bibinfo {author}
  {\bibfnamefont {R.}~\bibnamefont {Wiesendanger}}, \ and\ \bibinfo {author}
  {\bibfnamefont {S.}~\bibnamefont {Bl\"{u}gel}},\ }\href@noop {} {\bibfield
  {journal} {\bibinfo  {journal} {Phys. Rev. Lett.}\ }\textbf {\bibinfo
  {volume} {105}},\ \bibinfo {pages} {066601} (\bibinfo {year}
  {2010})}\BibitemShut {NoStop}%
\bibitem [{\citenamefont {Brede}\ \emph {et~al.}(2010)\citenamefont {Brede},
  \citenamefont {Atodiresei}, \citenamefont {Kuck}, \citenamefont
  {Lazi{\'{c}}}, \citenamefont {Caciuc}, \citenamefont {Morikawa},
  \citenamefont {Hoffmann}, \citenamefont {Bl{\"{u}}gel},\ and\ \citenamefont
  {Wiesendanger}}]{Brede10}%
  \BibitemOpen
  \bibfield  {author} {\bibinfo {author} {\bibfnamefont {J.}~\bibnamefont
  {Brede}}, \bibinfo {author} {\bibfnamefont {N.}~\bibnamefont {Atodiresei}},
  \bibinfo {author} {\bibfnamefont {S.}~\bibnamefont {Kuck}}, \bibinfo {author}
  {\bibfnamefont {P.}~\bibnamefont {Lazi{\'{c}}}}, \bibinfo {author}
  {\bibfnamefont {V.}~\bibnamefont {Caciuc}}, \bibinfo {author} {\bibfnamefont
  {Y.}~\bibnamefont {Morikawa}}, \bibinfo {author} {\bibfnamefont
  {G.}~\bibnamefont {Hoffmann}}, \bibinfo {author} {\bibfnamefont
  {S.}~\bibnamefont {Bl{\"{u}}gel}}, \ and\ \bibinfo {author} {\bibfnamefont
  {R.}~\bibnamefont {Wiesendanger}},\ }\href@noop {} {\bibfield  {journal}
  {\bibinfo  {journal} {Phys. Rev. Lett.}\ }\textbf {\bibinfo {volume} {105}},\
  \bibinfo {pages} {047204} (\bibinfo {year} {2010})}\BibitemShut {NoStop}%
\bibitem [{\citenamefont {Javaid}\ \emph {et~al.}(2010)\citenamefont {Javaid},
  \citenamefont {Bowen}, \citenamefont {Boukari}, \citenamefont {Joly},
  \citenamefont {Beaufrand}, \citenamefont {Chen}, \citenamefont {Dappe},
  \citenamefont {Schreurer}, \citenamefont {Kappler}, \citenamefont {Arabski},
  \citenamefont {Wulfhekel}, \citenamefont {Alouani},\ and\ \citenamefont
  {Beaurepaire}}]{Javaid10}%
  \BibitemOpen
  \bibfield  {author} {\bibinfo {author} {\bibfnamefont {S.}~\bibnamefont
  {Javaid}}, \bibinfo {author} {\bibfnamefont {M.}~\bibnamefont {Bowen}},
  \bibinfo {author} {\bibfnamefont {S.}~\bibnamefont {Boukari}}, \bibinfo
  {author} {\bibfnamefont {L.}~\bibnamefont {Joly}}, \bibinfo {author}
  {\bibfnamefont {J.}~\bibnamefont {Beaufrand}}, \bibinfo {author}
  {\bibfnamefont {X.}~\bibnamefont {Chen}}, \bibinfo {author} {\bibfnamefont
  {Y.~J.}\ \bibnamefont {Dappe}}, \bibinfo {author} {\bibfnamefont
  {F.}~\bibnamefont {Schreurer}}, \bibinfo {author} {\bibfnamefont {J.-P.}\
  \bibnamefont {Kappler}}, \bibinfo {author} {\bibfnamefont {J.}~\bibnamefont
  {Arabski}}, \bibinfo {author} {\bibfnamefont {W.}~\bibnamefont {Wulfhekel}},
  \bibinfo {author} {\bibfnamefont {M.}~\bibnamefont {Alouani}}, \ and\
  \bibinfo {author} {\bibfnamefont {E.}~\bibnamefont {Beaurepaire}},\
  }\href@noop {} {\bibfield  {journal} {\bibinfo  {journal} {Phys. Rev. Lett.}\
  }\textbf {\bibinfo {volume} {105}},\ \bibinfo {pages} {077201} (\bibinfo
  {year} {2010})}\BibitemShut {NoStop}%
\bibitem [{\citenamefont {Tran}\ \emph {et~al.}(2011)\citenamefont {Tran},
  \citenamefont {Wong}, \citenamefont {de~Jong}, \citenamefont {van~der Wiel},
  \citenamefont {Zhan},\ and\ \citenamefont {Fahlman}}]{Tran11}%
  \BibitemOpen
  \bibfield  {author} {\bibinfo {author} {\bibfnamefont {T.~L.~A.}\
  \bibnamefont {Tran}}, \bibinfo {author} {\bibfnamefont {P.~K.~J.}\
  \bibnamefont {Wong}}, \bibinfo {author} {\bibfnamefont {M.~P.}\ \bibnamefont
  {de~Jong}}, \bibinfo {author} {\bibfnamefont {W.~G.}\ \bibnamefont {van~der
  Wiel}}, \bibinfo {author} {\bibfnamefont {Y.~Q.}\ \bibnamefont {Zhan}}, \
  and\ \bibinfo {author} {\bibfnamefont {M.}~\bibnamefont {Fahlman}},\
  }\href@noop {} {\bibfield  {journal} {\bibinfo  {journal} {Appl. Phys.
  Lett.}\ }\textbf {\bibinfo {volume} {98}},\ \bibinfo {pages} {222505}
  (\bibinfo {year} {2011})}\BibitemShut {NoStop}%
\bibitem [{\citenamefont {Tran}\ \emph {et~al.}(2013)\citenamefont {Tran},
  \citenamefont {\c{C}ak{\i}r}, \citenamefont {Wong}, \citenamefont
  {Preobrajenski}, \citenamefont {Brocks}, \citenamefont {van~der Wiel},\ and\
  \citenamefont {de~Jong}}]{Tran13}%
  \BibitemOpen
  \bibfield  {author} {\bibinfo {author} {\bibfnamefont {T.~L.~A.}\
  \bibnamefont {Tran}}, \bibinfo {author} {\bibfnamefont {D.}~\bibnamefont
  {\c{C}ak{\i}r}}, \bibinfo {author} {\bibfnamefont {P.~K.~J.}\ \bibnamefont
  {Wong}}, \bibinfo {author} {\bibfnamefont {A.~B.}\ \bibnamefont
  {Preobrajenski}}, \bibinfo {author} {\bibfnamefont {G.}~\bibnamefont
  {Brocks}}, \bibinfo {author} {\bibfnamefont {W.~G.}\ \bibnamefont {van~der
  Wiel}}, \ and\ \bibinfo {author} {\bibfnamefont {M.~P.}\ \bibnamefont
  {de~Jong}},\ }\href@noop {} {\bibfield  {journal} {\bibinfo  {journal} {ACS
  Appl. Mater. interfaces}\ }\textbf {\bibinfo {volume} {5}},\ \bibinfo {pages}
  {837} (\bibinfo {year} {2013})}\BibitemShut {NoStop}%
\bibitem [{\citenamefont {Djeghloul}\ \emph {et~al.}(2013)\citenamefont
  {Djeghloul}, \citenamefont {Ibrahim}, \citenamefont {Cantoni}, \citenamefont
  {Bowen}, \citenamefont {Joly}, \citenamefont {Boukari}, \citenamefont
  {Ohresser}, \citenamefont {Bertran}, \citenamefont {Fevre}, \citenamefont
  {Thakur}, \citenamefont {Scheurer}, \citenamefont {Miyamachi}, \citenamefont
  {Mattana}, \citenamefont {Seneor}, \citenamefont {Jaafar}, \citenamefont
  {Rinaldi}, \citenamefont {Javaid}, \citenamefont {Arabski}, \citenamefont
  {Kappler}, \citenamefont {Wulfhekel}, \citenamefont {Brookes}, \citenamefont
  {Bertacco}, \citenamefont {Taleb-Ibrahimi}, \citenamefont {Alouani},
  \citenamefont {Beaurepaire},\ and\ \citenamefont {Weber}}]{Djeghloul13}%
  \BibitemOpen
  \bibfield  {author} {\bibinfo {author} {\bibfnamefont {F.}~\bibnamefont
  {Djeghloul}}, \bibinfo {author} {\bibfnamefont {F.}~\bibnamefont {Ibrahim}},
  \bibinfo {author} {\bibfnamefont {M.}~\bibnamefont {Cantoni}}, \bibinfo
  {author} {\bibfnamefont {M.}~\bibnamefont {Bowen}}, \bibinfo {author}
  {\bibfnamefont {L.}~\bibnamefont {Joly}}, \bibinfo {author} {\bibfnamefont
  {S.}~\bibnamefont {Boukari}}, \bibinfo {author} {\bibfnamefont
  {P.}~\bibnamefont {Ohresser}}, \bibinfo {author} {\bibfnamefont
  {F.}~\bibnamefont {Bertran}}, \bibinfo {author} {\bibfnamefont {P.~L.}\
  \bibnamefont {Fevre}}, \bibinfo {author} {\bibfnamefont {P.}~\bibnamefont
  {Thakur}}, \bibinfo {author} {\bibfnamefont {F.}~\bibnamefont {Scheurer}},
  \bibinfo {author} {\bibfnamefont {T.}~\bibnamefont {Miyamachi}}, \bibinfo
  {author} {\bibfnamefont {R.}~\bibnamefont {Mattana}}, \bibinfo {author}
  {\bibfnamefont {P.}~\bibnamefont {Seneor}}, \bibinfo {author} {\bibfnamefont
  {A.}~\bibnamefont {Jaafar}}, \bibinfo {author} {\bibfnamefont
  {C.}~\bibnamefont {Rinaldi}}, \bibinfo {author} {\bibfnamefont
  {S.}~\bibnamefont {Javaid}}, \bibinfo {author} {\bibfnamefont
  {J.}~\bibnamefont {Arabski}}, \bibinfo {author} {\bibfnamefont {J.-P.}\
  \bibnamefont {Kappler}}, \bibinfo {author} {\bibfnamefont {W.}~\bibnamefont
  {Wulfhekel}}, \bibinfo {author} {\bibfnamefont {N.~B.}\ \bibnamefont
  {Brookes}}, \bibinfo {author} {\bibfnamefont {R.}~\bibnamefont {Bertacco}},
  \bibinfo {author} {\bibfnamefont {A.}~\bibnamefont {Taleb-Ibrahimi}},
  \bibinfo {author} {\bibfnamefont {M.}~\bibnamefont {Alouani}}, \bibinfo
  {author} {\bibfnamefont {E.}~\bibnamefont {Beaurepaire}}, \ and\ \bibinfo
  {author} {\bibfnamefont {W.}~\bibnamefont {Weber}},\ }\href@noop {}
  {\bibfield  {journal} {\bibinfo  {journal} {Sci. Rep.}\ }\textbf {\bibinfo
  {volume} {3}},\ \bibinfo {pages} {1272} (\bibinfo {year} {2013})}\BibitemShut
  {NoStop}%
\bibitem [{\citenamefont {Schmaus}\ \emph {et~al.}(2011)\citenamefont
  {Schmaus}, \citenamefont {Bagrets}, \citenamefont {Nahas}, \citenamefont
  {Yamada}, \citenamefont {Bork}, \citenamefont {Bowen}, \citenamefont
  {Beaurepaire}, \citenamefont {Evers},\ and\ \citenamefont
  {Wulfhekel}}]{Schmaus11}%
  \BibitemOpen
  \bibfield  {author} {\bibinfo {author} {\bibfnamefont {S.}~\bibnamefont
  {Schmaus}}, \bibinfo {author} {\bibfnamefont {A.}~\bibnamefont {Bagrets}},
  \bibinfo {author} {\bibfnamefont {Y.}~\bibnamefont {Nahas}}, \bibinfo
  {author} {\bibfnamefont {T.~K.}\ \bibnamefont {Yamada}}, \bibinfo {author}
  {\bibfnamefont {A.}~\bibnamefont {Bork}}, \bibinfo {author} {\bibfnamefont
  {M.}~\bibnamefont {Bowen}}, \bibinfo {author} {\bibfnamefont
  {E.}~\bibnamefont {Beaurepaire}}, \bibinfo {author} {\bibfnamefont
  {F.}~\bibnamefont {Evers}}, \ and\ \bibinfo {author} {\bibfnamefont
  {W.}~\bibnamefont {Wulfhekel}},\ }\href@noop {} {\bibfield  {journal}
  {\bibinfo  {journal} {Nature Nanotechnol.}\ }\textbf {\bibinfo {volume}
  {6}},\ \bibinfo {pages} {185} (\bibinfo {year} {2011})}\BibitemShut {NoStop}%
\bibitem [{\citenamefont {Miyamachi}\ \emph {et~al.}(2012)\citenamefont
  {Miyamachi}, \citenamefont {Gruber}, \citenamefont {Davesne}, \citenamefont
  {Bowen}, \citenamefont {Boukari}, \citenamefont {Joly}, \citenamefont
  {Scheurer}, \citenamefont {Rogez}, \citenamefont {Yamada}, \citenamefont
  {Ohresser}, \citenamefont {Beaurepaire},\ and\ \citenamefont
  {Wulfhekel}}]{Miyamachi12}%
  \BibitemOpen
  \bibfield  {author} {\bibinfo {author} {\bibfnamefont {T.}~\bibnamefont
  {Miyamachi}}, \bibinfo {author} {\bibfnamefont {M.}~\bibnamefont {Gruber}},
  \bibinfo {author} {\bibfnamefont {V.}~\bibnamefont {Davesne}}, \bibinfo
  {author} {\bibfnamefont {M.}~\bibnamefont {Bowen}}, \bibinfo {author}
  {\bibfnamefont {S.}~\bibnamefont {Boukari}}, \bibinfo {author} {\bibfnamefont
  {L.}~\bibnamefont {Joly}}, \bibinfo {author} {\bibfnamefont {F.}~\bibnamefont
  {Scheurer}}, \bibinfo {author} {\bibfnamefont {G.}~\bibnamefont {Rogez}},
  \bibinfo {author} {\bibfnamefont {T.~K.}\ \bibnamefont {Yamada}}, \bibinfo
  {author} {\bibfnamefont {P.}~\bibnamefont {Ohresser}}, \bibinfo {author}
  {\bibfnamefont {E.}~\bibnamefont {Beaurepaire}}, \ and\ \bibinfo {author}
  {\bibfnamefont {W.}~\bibnamefont {Wulfhekel}},\ }\href@noop {} {\bibfield
  {journal} {\bibinfo  {journal} {Nature Commun.}\ }\textbf {\bibinfo {volume}
  {3}},\ \bibinfo {pages} {938} (\bibinfo {year} {2012})}\BibitemShut {NoStop}%
\bibitem [{\citenamefont {Rocha}\ \emph {et~al.}(2005)\citenamefont {Rocha},
  \citenamefont {Garcia-Suarez}, \citenamefont {Bailey}, \citenamefont
  {Lambert}, \citenamefont {Ferrer},\ and\ \citenamefont {Sanvito}}]{Rocha05}%
  \BibitemOpen
  \bibfield  {author} {\bibinfo {author} {\bibfnamefont {A.~R.}\ \bibnamefont
  {Rocha}}, \bibinfo {author} {\bibfnamefont {V.~M.}\ \bibnamefont
  {Garcia-Suarez}}, \bibinfo {author} {\bibfnamefont {S.}~\bibnamefont
  {Bailey}}, \bibinfo {author} {\bibfnamefont {C.}~\bibnamefont {Lambert}},
  \bibinfo {author} {\bibfnamefont {J.}~\bibnamefont {Ferrer}}, \ and\ \bibinfo
  {author} {\bibfnamefont {S.}~\bibnamefont {Sanvito}},\ }\href@noop {}
  {\bibfield  {journal} {\bibinfo  {journal} {Nature Mater.}\ }\textbf
  {\bibinfo {volume} {4}},\ \bibinfo {pages} {335} (\bibinfo {year}
  {2005})}\BibitemShut {NoStop}%
\bibitem [{\citenamefont {Ning}\ \emph {et~al.}(2008)\citenamefont {Ning},
  \citenamefont {Zhu}, \citenamefont {Wang},\ and\ \citenamefont
  {Guo}}]{Ning08}%
  \BibitemOpen
  \bibfield  {author} {\bibinfo {author} {\bibfnamefont {Z.}~\bibnamefont
  {Ning}}, \bibinfo {author} {\bibfnamefont {Y.}~\bibnamefont {Zhu}}, \bibinfo
  {author} {\bibfnamefont {J.}~\bibnamefont {Wang}}, \ and\ \bibinfo {author}
  {\bibfnamefont {H.}~\bibnamefont {Guo}},\ }\href@noop {} {\bibfield
  {journal} {\bibinfo  {journal} {Phys. Rev. Lett.}\ }\textbf {\bibinfo
  {volume} {100}},\ \bibinfo {pages} {056803} (\bibinfo {year}
  {2008})}\BibitemShut {NoStop}%
\bibitem [{\citenamefont {Koleini}\ and\ \citenamefont
  {Brandbyge}(2012)}]{Koleini12}%
  \BibitemOpen
  \bibfield  {author} {\bibinfo {author} {\bibfnamefont {M.}~\bibnamefont
  {Koleini}}\ and\ \bibinfo {author} {\bibfnamefont {M.}~\bibnamefont
  {Brandbyge}},\ }\href@noop {} {\bibfield  {journal} {\bibinfo  {journal}
  {Beilstein J. Nanotechnol.}\ }\textbf {\bibinfo {volume} {3}},\ \bibinfo
  {pages} {589} (\bibinfo {year} {2012})}\BibitemShut {NoStop}%
\bibitem [{\citenamefont {Momma}\ and\ \citenamefont {Izumi}(2011)}]{VESTA}%
  \BibitemOpen
  \bibfield  {author} {\bibinfo {author} {\bibfnamefont {K.}~\bibnamefont
  {Momma}}\ and\ \bibinfo {author} {\bibfnamefont {F.}~\bibnamefont {Izumi}},\
  }\href@noop {} {\bibfield  {journal} {\bibinfo  {journal} {J. Appl.
  Crystallogr.}\ }\textbf {\bibinfo {volume} {44}},\ \bibinfo {pages} {1272}
  (\bibinfo {year} {2011})}\BibitemShut {NoStop}%
\bibitem [{foo({\natexlab{a}})}]{footnote:MR}%
  \BibitemOpen
  \bibinfo {note} {We use the conventional
  definition of the magnetoresistance $\mathrm{MR} = (I_\mathrm{P} -
  I_\mathrm{AP})/I_\mathrm{AP}$, where $I_\mathrm{P(AP)}$ is the total current
  with the magnetizations of the two ferromagnetic metal electrodes parallel
  (anti-parallel).}\BibitemShut {Stop}%
\bibitem [{\citenamefont {Landauer}(1957)}]{Landauer57}%
  \BibitemOpen
  \bibfield  {author} {\bibinfo {author} {\bibfnamefont {R.}~\bibnamefont
  {Landauer}},\ }\href@noop {} {\bibfield  {journal} {\bibinfo  {journal} {IBM
  J. Res. Dev.}\ }\textbf {\bibinfo {volume} {1}},\ \bibinfo {pages} {223}
  (\bibinfo {year} {1957})}\BibitemShut {NoStop}%
\bibitem [{\citenamefont {Brandbyge}\ \emph {et~al.}(2002)\citenamefont
  {Brandbyge}, \citenamefont {Mozos}, \citenamefont {Ordej\'{o}n},
  \citenamefont {Taylor},\ and\ \citenamefont {Stokbro}}]{Transiesta02}%
  \BibitemOpen
  \bibfield  {author} {\bibinfo {author} {\bibfnamefont {M.}~\bibnamefont
  {Brandbyge}}, \bibinfo {author} {\bibfnamefont {J.-L.}\ \bibnamefont
  {Mozos}}, \bibinfo {author} {\bibfnamefont {P.}~\bibnamefont {Ordej\'{o}n}},
  \bibinfo {author} {\bibfnamefont {J.}~\bibnamefont {Taylor}}, \ and\ \bibinfo
  {author} {\bibfnamefont {K.}~\bibnamefont {Stokbro}},\ }\href@noop {}
  {\bibfield  {journal} {\bibinfo  {journal} {Phys. Rev. B}\ }\textbf {\bibinfo
  {volume} {65}},\ \bibinfo {pages} {165401} (\bibinfo {year}
  {2002})}\BibitemShut {NoStop}%
\bibitem [{\citenamefont {Khomyakov}\ \emph {et~al.}(2005)\citenamefont
  {Khomyakov}, \citenamefont {Brocks}, \citenamefont {Karpan}, \citenamefont
  {Zwierzycki},\ and\ \citenamefont {Kelly}}]{Khomyakov05}%
  \BibitemOpen
  \bibfield  {author} {\bibinfo {author} {\bibfnamefont {P.~A.}\ \bibnamefont
  {Khomyakov}}, \bibinfo {author} {\bibfnamefont {G.}~\bibnamefont {Brocks}},
  \bibinfo {author} {\bibfnamefont {V.~M.}\ \bibnamefont {Karpan}}, \bibinfo
  {author} {\bibfnamefont {M.}~\bibnamefont {Zwierzycki}}, \ and\ \bibinfo
  {author} {\bibfnamefont {P.~J.}\ \bibnamefont {Kelly}},\ }\href@noop {}
  {\bibfield  {journal} {\bibinfo  {journal} {Phys. Rev. B}\ }\textbf {\bibinfo
  {volume} {72}},\ \bibinfo {pages} {035450} (\bibinfo {year}
  {2005})}\BibitemShut {NoStop}%
\bibitem [{\citenamefont {Mingo}\ \emph {et~al.}(1996)\citenamefont {Mingo},
  \citenamefont {Jurczyszyn}, \citenamefont {Garcia-Vidal}, \citenamefont
  {Saiz-Pardo}, \citenamefont {de~Andres}, \citenamefont {Flores},
  \citenamefont {Wu},\ and\ \citenamefont {More}}]{Mingo96}%
  \BibitemOpen
  \bibfield  {author} {\bibinfo {author} {\bibfnamefont {N.}~\bibnamefont
  {Mingo}}, \bibinfo {author} {\bibfnamefont {L.}~\bibnamefont {Jurczyszyn}},
  \bibinfo {author} {\bibfnamefont {F.~J.}\ \bibnamefont {Garcia-Vidal}},
  \bibinfo {author} {\bibfnamefont {R.}~\bibnamefont {Saiz-Pardo}}, \bibinfo
  {author} {\bibfnamefont {P.~L.}\ \bibnamefont {de~Andres}}, \bibinfo {author}
  {\bibfnamefont {F.}~\bibnamefont {Flores}}, \bibinfo {author} {\bibfnamefont
  {S.~Y.}\ \bibnamefont {Wu}}, \ and\ \bibinfo {author} {\bibfnamefont
  {W.}~\bibnamefont {More}},\ }\href@noop {} {\bibfield  {journal} {\bibinfo
  {journal} {Phys. Rev. B}\ }\textbf {\bibinfo {volume} {54}},\ \bibinfo
  {pages} {2225} (\bibinfo {year} {1996})}\BibitemShut {NoStop}%
\bibitem [{\citenamefont {Mathon}(1997)}]{Mathon97}%
  \BibitemOpen
  \bibfield  {author} {\bibinfo {author} {\bibfnamefont {J.}~\bibnamefont
  {Mathon}},\ }\href@noop {} {\bibfield  {journal} {\bibinfo  {journal} {Phys.
  Rev. B}\ }\textbf {\bibinfo {volume} {56}},\ \bibinfo {pages} {11810}
  (\bibinfo {year} {1997})}\BibitemShut {NoStop}%
\bibitem [{\citenamefont {Meunier}\ and\ \citenamefont
  {Lambin}(1998)}]{Meunier98}%
  \BibitemOpen
  \bibfield  {author} {\bibinfo {author} {\bibfnamefont {V.}~\bibnamefont
  {Meunier}}\ and\ \bibinfo {author} {\bibfnamefont {P.}~\bibnamefont
  {Lambin}},\ }\href@noop {} {\bibfield  {journal} {\bibinfo  {journal} {Phys.
  Rev. Lett.}\ }\textbf {\bibinfo {volume} {81}},\ \bibinfo {pages} {5588}
  (\bibinfo {year} {1998})}\BibitemShut {NoStop}%
\bibitem [{\citenamefont {Julli{\`{e}}re}(1975)}]{Julliere75}%
  \BibitemOpen
  \bibfield  {author} {\bibinfo {author} {\bibfnamefont {M.}~\bibnamefont
  {Julli{\`{e}}re}},\ }\href@noop {} {\bibfield  {journal} {\bibinfo  {journal}
  {Phys. Lett. A}\ }\textbf {\bibinfo {volume} {54}},\ \bibinfo {pages} {225}
  (\bibinfo {year} {1975})}\BibitemShut {NoStop}%
\bibitem [{\citenamefont {Moodera}\ and\ \citenamefont
  {Mathon}(1999)}]{Moodera99}%
  \BibitemOpen
  \bibfield  {author} {\bibinfo {author} {\bibfnamefont {J.~S.}\ \bibnamefont
  {Moodera}}\ and\ \bibinfo {author} {\bibfnamefont {J.}~\bibnamefont
  {Mathon}},\ }\href@noop {} {\bibfield  {journal} {\bibinfo  {journal} {J.
  Magn. Magn. Mater.}\ }\textbf {\bibinfo {volume} {200}},\ \bibinfo {pages}
  {248} (\bibinfo {year} {1999})}\BibitemShut {NoStop}%
\bibitem [{\citenamefont {Belashchenko}\ \emph {et~al.}(2004)\citenamefont
  {Belashchenko}, \citenamefont {Tsymbal}, \citenamefont {van Schilfgaarde},
  \citenamefont {Stewart}, \citenamefont {Oleynik},\ and\ \citenamefont
  {Jaswal}}]{Belashchenko04}%
  \BibitemOpen
  \bibfield  {author} {\bibinfo {author} {\bibfnamefont {K.~D.}\ \bibnamefont
  {Belashchenko}}, \bibinfo {author} {\bibfnamefont {E.~Y.}\ \bibnamefont
  {Tsymbal}}, \bibinfo {author} {\bibfnamefont {M.}~\bibnamefont {van
  Schilfgaarde}}, \bibinfo {author} {\bibfnamefont {D.~A.}\ \bibnamefont
  {Stewart}}, \bibinfo {author} {\bibfnamefont {I.~I.}\ \bibnamefont
  {Oleynik}}, \ and\ \bibinfo {author} {\bibfnamefont {S.~S.}\ \bibnamefont
  {Jaswal}},\ }\href@noop {} {\bibfield  {journal} {\bibinfo  {journal} {Phys.
  Rev. B}\ }\textbf {\bibinfo {volume} {69}},\ \bibinfo {pages} {174408}
  (\bibinfo {year} {2004})}\BibitemShut {NoStop}%
\bibitem [{\citenamefont {Xu}\ \emph {et~al.}(2006)\citenamefont {Xu},
  \citenamefont {Karpan}, \citenamefont {Xia}, \citenamefont {Zwierzycki},
  \citenamefont {Marushchenko},\ and\ \citenamefont {Kelly}}]{Xu06}%
  \BibitemOpen
  \bibfield  {author} {\bibinfo {author} {\bibfnamefont {P.~X.}\ \bibnamefont
  {Xu}}, \bibinfo {author} {\bibfnamefont {V.~M.}\ \bibnamefont {Karpan}},
  \bibinfo {author} {\bibfnamefont {K.}~\bibnamefont {Xia}}, \bibinfo {author}
  {\bibfnamefont {M.}~\bibnamefont {Zwierzycki}}, \bibinfo {author}
  {\bibfnamefont {I.}~\bibnamefont {Marushchenko}}, \ and\ \bibinfo {author}
  {\bibfnamefont {P.~J.}\ \bibnamefont {Kelly}},\ }\href@noop {} {\bibfield
  {journal} {\bibinfo  {journal} {Phys. Rev. B}\ }\textbf {\bibinfo {volume}
  {73}},\ \bibinfo {pages} {180402} (\bibinfo {year} {2006})}\BibitemShut
  {NoStop}%
\bibitem [{\citenamefont {Kresse}\ and\ \citenamefont {Hafner}(1993)}]{vasp-1}%
  \BibitemOpen
  \bibfield  {author} {\bibinfo {author} {\bibfnamefont {G.}~\bibnamefont
  {Kresse}}\ and\ \bibinfo {author} {\bibfnamefont {J.}~\bibnamefont
  {Hafner}},\ }\href@noop {} {\bibfield  {journal} {\bibinfo  {journal} {Phys.
  Rev. B}\ }\textbf {\bibinfo {volume} {47}},\ \bibinfo {pages} {558} (\bibinfo
  {year} {1993})}\BibitemShut {NoStop}%
\bibitem [{\citenamefont {Kresse}\ and\ \citenamefont
  {Furthm\"{u}ller}(1996)}]{vasp-2}%
  \BibitemOpen
  \bibfield  {author} {\bibinfo {author} {\bibfnamefont {G.}~\bibnamefont
  {Kresse}}\ and\ \bibinfo {author} {\bibfnamefont {J.}~\bibnamefont
  {Furthm\"{u}ller}},\ }\href@noop {} {\bibfield  {journal} {\bibinfo
  {journal} {Phys. Rev. B}\ }\textbf {\bibinfo {volume} {54}},\ \bibinfo
  {pages} {11169} (\bibinfo {year} {1996})}\BibitemShut {NoStop}%
\bibitem [{\citenamefont {Verheijen}\ \emph {et~al.}(1992)\citenamefont
  {Verheijen}, \citenamefont {Meekes}, \citenamefont {Meijer}, \citenamefont
  {Bennema}, \citenamefont {de~Boer}, \citenamefont {van Smaalen},
  \citenamefont {van Tendeloo}, \citenamefont {Amelinckx}, \citenamefont
  {Muto},\ and\ \citenamefont {van Landuyt}}]{Verheijen:cp92}%
  \BibitemOpen
  \bibfield  {author} {\bibinfo {author} {\bibfnamefont {M.}~\bibnamefont
  {Verheijen}}, \bibinfo {author} {\bibfnamefont {H.}~\bibnamefont {Meekes}},
  \bibinfo {author} {\bibfnamefont {G.}~\bibnamefont {Meijer}}, \bibinfo
  {author} {\bibfnamefont {P.}~\bibnamefont {Bennema}}, \bibinfo {author}
  {\bibfnamefont {J.~L.}\ \bibnamefont {de~Boer}}, \bibinfo {author}
  {\bibfnamefont {S.}~\bibnamefont {van Smaalen}}, \bibinfo {author}
  {\bibfnamefont {G.}~\bibnamefont {van Tendeloo}}, \bibinfo {author}
  {\bibfnamefont {S.}~\bibnamefont {Amelinckx}}, \bibinfo {author}
  {\bibfnamefont {S.}~\bibnamefont {Muto}}, \ and\ \bibinfo {author}
  {\bibfnamefont {J.}~\bibnamefont {van Landuyt}},\ }\href@noop {} {\bibfield
  {journal} {\bibinfo  {journal} {Chem. Phys.}\ }\textbf {\bibinfo {volume}
  {166}},\ \bibinfo {pages} {287} (\bibinfo {year} {1992})}\BibitemShut
  {NoStop}%
\bibitem [{\citenamefont {Soler}\ \emph {et~al.}(2002)\citenamefont {Soler},
  \citenamefont {Artacho}, \citenamefont {Gale}, \citenamefont {Garc\'{i}a},
  \citenamefont {Junquera}, \citenamefont {Ordej\'{o}n},\ and\ \citenamefont
  {S\'{a}nchez-Portal}}]{siesta}%
  \BibitemOpen
  \bibfield  {author} {\bibinfo {author} {\bibfnamefont {J.~M.}\ \bibnamefont
  {Soler}}, \bibinfo {author} {\bibfnamefont {E.}~\bibnamefont {Artacho}},
  \bibinfo {author} {\bibfnamefont {J.~D.}\ \bibnamefont {Gale}}, \bibinfo
  {author} {\bibfnamefont {A.}~\bibnamefont {Garc\'{i}a}}, \bibinfo {author}
  {\bibfnamefont {J.}~\bibnamefont {Junquera}}, \bibinfo {author}
  {\bibfnamefont {P.}~\bibnamefont {Ordej\'{o}n}}, \ and\ \bibinfo {author}
  {\bibfnamefont {D.}~\bibnamefont {S\'{a}nchez-Portal}},\ }\href@noop {}
  {\bibfield  {journal} {\bibinfo  {journal} {J. Phys.: Condens. Matter}\
  }\textbf {\bibinfo {volume} {14}},\ \bibinfo {pages} {2745} (\bibinfo {year}
  {2002})}\BibitemShut {NoStop}%
\bibitem [{\citenamefont {Troullier}\ and\ \citenamefont {Martins}(1991)}]{tm}%
  \BibitemOpen
  \bibfield  {author} {\bibinfo {author} {\bibfnamefont {N.}~\bibnamefont
  {Troullier}}\ and\ \bibinfo {author} {\bibfnamefont {J.~L.}\ \bibnamefont
  {Martins}},\ }\href@noop {} {\bibfield  {journal} {\bibinfo  {journal} {Phys.
  Rev. B}\ }\textbf {\bibinfo {volume} {43}},\ \bibinfo {pages} {1993}
  (\bibinfo {year} {1991})}\BibitemShut {NoStop}%
\bibitem [{foo({\natexlab{b}})}]{footnote:SIESTA}%
  \BibitemOpen
  \bibinfo {note} {SIESTA gives a somewhat
  larger exchange splitting than VASP, which leads to 3.7, 3.0\% larger
  magnetic moments on the atoms in bulk Fe and at the Fe(001) surface,
  respectively. This difference can be traced back to the use of
  pseudo-potentials (SIESTA), instead of an all-electron scheme
  (VASP).}\BibitemShut {Stop}%
\bibitem [{\citenamefont {Stroscio}\ \emph {et~al.}(1995)\citenamefont
  {Stroscio}, \citenamefont {Pierce}, \citenamefont {Davies},\ and\
  \citenamefont {Celotta}}]{Stroscio95}%
  \BibitemOpen
  \bibfield  {author} {\bibinfo {author} {\bibfnamefont {J.~A.}\ \bibnamefont
  {Stroscio}}, \bibinfo {author} {\bibfnamefont {D.~T.}\ \bibnamefont
  {Pierce}}, \bibinfo {author} {\bibfnamefont {A.}~\bibnamefont {Davies}}, \
  and\ \bibinfo {author} {\bibfnamefont {R.~J.}\ \bibnamefont {Celotta}},\
  }\href@noop {} {\bibfield  {journal} {\bibinfo  {journal} {Phys. Rev. Lett.}\
  }\textbf {\bibinfo {volume} {75}},\ \bibinfo {pages} {2960} (\bibinfo {year}
  {1995})}\BibitemShut {NoStop}%
\end{thebibliography}
%

\end{document}